\documentclass[preprint2]{aastex}

\usepackage{graphicx,epsfig,amssymb,layout,verbatim,rotating,calc,mathrsfs}
\usepackage[sumlimits,intlimits,namelimits]{amsmath}
\usepackage{graphics}
\usepackage{rotate}
\PassOptionsToPackage{linktocpage}{hyperref}
\usepackage{natbib}

\def\apjl{ApJL}

\def\apjs{ApJSS}

\def\gsim{\mathrel{\raise.5ex\hbox{$>$}\mkern-14mu
             \lower0.6ex\hbox{$\sim$}}}
\def\lsim{\mathrel{\raise.3ex\hbox{$<$}\mkern-14mu
             \lower0.6ex\hbox{$\sim$}}}

\shorttitle{Toward a Realistic Pulsar Magnetosphere}
\shortauthors{Kalapotharakos et al.}

\begin{document}


\title{Toward a Realistic Pulsar Magnetosphere}


\author{Constantinos Kalapotharakos\altaffilmark{1,2}, Demosthenes Kazanas\altaffilmark{2}, Alice Harding\altaffilmark{2} and Ioannis Contopoulos\altaffilmark{3}}
\affil{$^1$University of Maryland, College Park (UMDCP/CRESST), College Park, MD 20742, USA;\\
$^2$Astrophysics Science Division, NASA/Goddard Space Flight Center, Greenbelt, MD 20771, USA;\\
$^3$Research Center for Astronomy and Applied Mathematics, Academy
of Athens, Athens 11527, Greece}
\email{constantinos.kalapotharakos@nasa.gov}



\begin{abstract}
We present the magnetic and electric field structures as well as the
currents and charge densities of pulsar magnetospheres which do not
obey the ideal condition, ${\bf E \cdot B =0}$. Since the
acceleration of particles and the production of radiation requires
the presence of an electric field component parallel to the magnetic
field, ${\bf E}_\parallel$, the structure of non-Ideal pulsar
magnetospheres is intimately related to the production of pulsar
radiation. Therefore, knowledge of the structure of non-Ideal pulsar
magnetospheres is important because their comparison (including
models for the production of radiation) with observations will
delineate the physics and the parameters underlying the pulsar
radiation problem. We implement a variety of prescriptions that
support nonzero values for ${\bf E}_\parallel$ and explore their
effects on the structure of the resulting magnetospheres. We produce
families of solutions that span the entire range between the vacuum
and the (ideal) Force-Free Electrodynamic solutions. We also compute
the amount of dissipation as a fraction of the Poynting flux for
pulsars of different angles between the rotation and magnetic axes
and conclude that this is at most 20-40\% (depending on the
non-ideal prescription) in the aligned rotator and 10\% in the
perpendicular one. We present also the limiting solutions with the
property $J=\rho c$ and discuss their possible implication on the
determination of the ``on/off'' states of the intermittent pulsars.
Finally, we find that solutions with values of $J$ greater than
those needed to null ${\bf E}_\parallel$ locally produce
oscillations, potentially observable in the data.
\end{abstract}


\section{Introduction}

Pulsars are extraordinary objects powered by magnetic fields of
order $10^{12}$ G anchored onto neutron stars rotating with periods
$\sim 10^{-3} - 10$ s. These fields mediate the conversion of their
rotational energy into MHD winds and the acceleration of particles
to energies sufficiently high to produce GeV photons. Their
electromagnetic emission is quite complex and ranges from the radio
to the multi-GeV $\gamma-$ray regime.

Pulsar radiation has been the subject of many studies since their
discovery; however, despite such a multi-decade effort by a large
number of researchers its details are not fully understood. There
are several reasons for this. On the observational side, the
sensitivity of the $\gamma-$ ray instruments (the band where most of
the pulsar radiation power is emitted) has become adequate for the
determination of a significant number of pulsar spectra only
recently.
Since its launch in 2008, the {\it Fermi Gamma-Ray Space
Telescope} has revolutionized pulsar physics by detecting nearly 90
$\gamma-$ray pulsars so far, with a variety of well-measured light
curves. On the theoretical side, the structure of the pulsar
magnetosphere, even for the simplest axisymmetric case, remained
uncertain for a long time. The modeling of pulsar magnetospheres has
seen rapid advancements only very recently.

The original treatment of the non-axisymmetric rotating stellar
dipole in vacuum \citep{deutsch1955} established the presence of an
electric field component, $ {\bf E}_{\vert \vert}$\footnote{ ${\bf
E}_{\parallel}\equiv ({\bf E}\cdot {\bf B}){\bf B}/B^2$, where ${\bf
E, B}$ are the electric and magnetic field vectors.}, parallel to
the magnetic field on the star's surface. Then \citet{GJ69}
suggested that this component of the electric field would pull
charges out of the pulsar, opening up magnetic field lines that
cross the Light Cylinder (hereafter LC), to produce an MHD wind with
currents flowing out of and into the neutron star polar cap. The
charge density needed to achieve this is known as the Goldreich -
Julian density $\rho_{\rm GJ}$; an associated current density
$J_{\rm GJ} = c \rho_{\rm GJ} $ is obtained if we allow these
charges to move outward at nearly the speed of light.

Following the work of \citet{GJ69}, a number of attempts were made
to produce the structure of the pulsar magnetosphere. Thus,
\cite{SW73} were able to reduce the structure of the axisymmetric
magnetosphere to a single equation for the poloidal magnetic flux,
the so-called pulsar equation. However an exact solution of this
equation that would produce the magnetic field structure across the
LC was missing for almost 3 decades \citep[for a review of the
problems and the various attempts to produce a consistent pulsar
magnetosphere, even in the axisymmetric case, see][]{Michel82}. In
fact, the inability of models to determine the magnetospheric
structure across the LC led people to speculate that its structure
is not smooth across this surface and assumed that the observed
radiation is the result of magnetic field discontinuities across the
LC \citep{MS94}.

The structure of the axisymmetric pulsar magnetosphere within Force
Free Electrodynamics (hereafter FFE) (i.e. assuming that the only
surviving component of the electric field is that perpendicular to
the magnetic field, or equivalently that ${\bf E \cdot B =0}$) was
given by \citet*{ckf99} (hereafter CKF) who used an iterative
procedure to determine the current distribution on the LC so that
the magnetic field lines should cross this surface smoothly. Besides
allowing a smooth transition of the magnetic field lines across the
LC, this solution also resolved the issue of the current closure in
pulsar magnetospheres: The current starts near the polar axis and
closes, for $R > R_{\rm LC}$ (here $R_{\rm LC}$ is the radius of the
LC), mainly along an equatorial current sheet (allowing also a small
amount of return current at a finite height above it). For  $R <
R_{\rm LC}$ the flow is mainly along the separatrix of the open and
closed field lines. The axisymmetric solution has since been
confirmed and further studied by several others
\citep{gruzinov2005,timokhin2006,komissarov2006,mckinney2006}, in
particular with respect to its properties near the Y-point
\citep[the intersection of the last closed field line and the
LC;][]{uzd2003}.

The first non-axisymmetric (3D), oblique rotator magnetosphere was
presented by \citet{S2006}, who used a time-dependent numerical code
to advance the magnetic and electric fields under FFE conditions to
steady state. These simulations confirmed the general picture of
current closure established by the CKF solution and produced a
structure very similar to that of CKF in the axisymmetric case.
Similar simulations were performed by \citet{kc2009} who, using a
Perfectly Matched Layer (hereafter PML) at the outer boundary of
their computational domain, were able to follow these simulations
over many stellar rotation periods. In general, the 3D
magnetosphere, just like the axisymmetric one, consists of regions
of closed and open field lines with a large scale electric current
circuit established along open magnetic field lines. In the 3D case,
the current sheet needed for the global current closure is in fact
undulating, as foreseen in the kinematic solution of \citet{Bogo99}.
As shown by recent simulations, the undulating current sheet
structure is stable at least to distances of $10 R_{LC}$ and for
several stellar rotation periods \citep*{kalconkaz2011}.

The FFE solutions presented so far, by construction, do not allow
any electric field component parallel to the magnetic field ${\bf
B}$ (${\bf E \cdot B =0}$); as such they do not provide the
possibility of particle acceleration and the production of
radiation, in disagreement with observation. Nonetheless, models of
pulsar radiation have been constructed assuming either a vacuum
dipole geometry, or the field line geometry of the FFE
magnetosphere. Polar cap particle acceleration, $\gamma-$ray
emission and the formation of pair cascades were discussed over many
years as a means of accounting for the observed high-energy
radiation \citep{rudsut1975,dauhar1996}. However, measurements by
{\it Fermi} of the cutoff shape of the Vela pulsar spectrum
\citep{abdoetal2009a} has ruled out the super-exponential shape of
the spectrum predicted by polar cap models due to the attenuation of
pair production. Models placing the origin of the high-energy
emission in the outer magnetosphere or beyond are now universally
favored. Slot gap models \citep{mushar2004,hsdf2008}, with
acceleration and emission in a narrow layer along the last open
field line from the neutron star surface to near the LC, and outer
gap models \citep{romyad1995,romani1996,takchache2007} produce light
curves and spectra in first-order agreement with the {\it Fermi}
data \citep*{romwat2010,venhargui2009}. All of the above models
assume a retarded vacuum magnetic field. There are also kinematic
models of pulsar emission derived by injecting photons along the
magnetic field lines of either the FFE magnetosphere \citep{BS10a}
or those of the vacuum field geometry \citep{BS10b}, or along the
electric current near the equatorial current sheet where the ratio
$J/\rho c \rightarrow 1$ in FFE \citep{ck2010}. These models also
produce light curves in broad agreement with observations. However,
fits of {\it Fermi} data light curves with slot gap and outer gap
models in FFE geometry are less favorable than those for the vacuum
dipole geometry \citep{hardingetal2011}.

While FFE magnetospheres do not allow for particle acceleration,
they do, however, provide a global magnetospheric structure
consistent with the boundary conditions on the neutron star surface
(a perfect conductor), a smooth transition through the LC and the
establishment of a large scale MHD wind. They also determine {\em
both} the currents {\em and} the sign of the charges that flow in
the magnetosphere. However, they do not provide any information
about the particle production that might be necessary to support the
underlying charge conservation and electric current continuity. This
was apparent already in the CKF solution and was also noted in
\citet{Michel82}: There are regions, most notably the intersections
of the null charge surface with open magnetic field lines and the
region near the neutral (Y) point of the LC where the FFE charge
density changes sign along an open field line (along which particles
are supposed to flow freely). The numerical FFE solution has no
problem in these regions because it can simply supply at will the
charges necessary for current closure and for consistency with the
global boundary conditions. This, however, requires sources of
electron-positron pairs at the proper regions and of the proper
strength. Even though the FFE treatment of the pulsar magnetosphere
involves no such processes, these must be taken into account
self-consistently. Viewed this way, pulsar radiation is the outcome
of the ``tension" between the boundary conditions imposed by the
global magnetospheric structure and the particle production
necessary for charge conservation and electric current continuity.

\begin{figure*}
  \centering
  \includegraphics[width=\textwidth]{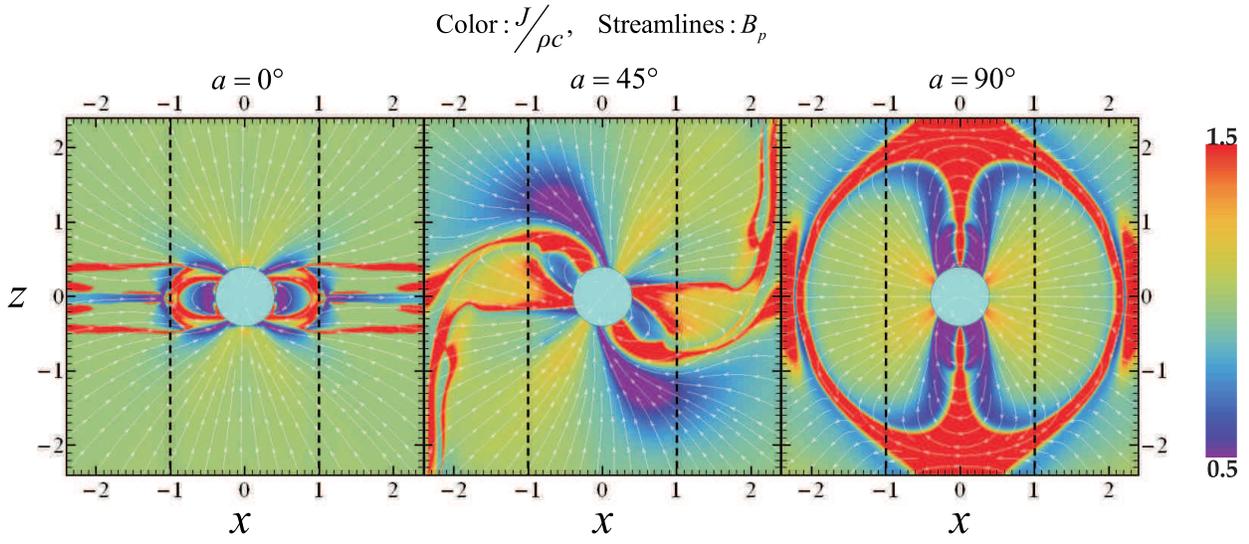}\\
  \caption{The FFE solutions for the inclination angles $a$ indicated in the figure.
 Each panel shows the poloidal magnetic field lines (white colored lines) and the ratio $J/\rho c$
 in the color scale shown in the figure. We note that the color scale saturates below the
 value 0.5 and above 1.5 so that details around the value 1 become evident.
 The dashed vertical lines denote the Light Cylinder (LC). The length unit is equal to $R_{\rm LC}$.
 We observe that the magnetic field lines open beyond the LC. The fraction of the polar cap region
 with space-like currents ($J/\rho c\geq 1$) increases with $a$ so that, while in the aligned rotator
 almost the entire polar cap is filled with time-like currents ($J/\rho c\leq 1$), in the
 perpendicular rotator the entire polar cap is filled with space-like currents ($J/\rho c\geq
 1$).}
 \label{fig00}
\end{figure*}

In Figs.~\ref{fig00}a,b,c we present the structure of the FFE
magnetosphere on the poloidal plane $\boldsymbol{\mu} -
\mathbf{\Omega}$ for inclination angles $a = (0^\circ, 45^\circ,
90^\circ)$. These figures depict the poloidal magnetic field lines
and in color the value of $J/\rho c$.\footnote{Hereafter, wherever
we use the ratio $J/\rho c$ we consider its absolute value.} The
vertical dashed lines denote the position of the LC. In the
axisymmetric case the value of $J/\rho c$ is less than one in most
space and exceeds this value only in the null surfaces and the
separatrix between open and closed field lines. The value 1 of this
ratio is significant because it denotes the maximum current density
for carriers of single sign charge. Values of this ratio greater
than 1, by necessity denote the presence of both positive and
negative charges moving in opposite directions. For $a \simeq 0$,
$J/\rho c < 1$ in most of the polar cap region, however, as the
value of inclination $a$ increases so does the size of regions with
$J/\rho c > 1$; in the perpendicular rotator we see clearly that
$J/\rho c > 1$ over the entire polar cap region.

With the above considerations in mind, the question that immediately
arises is how would the structure of the pulsar magnetosphere and
the consequent production of radiation be modified, if the ideal
condition, namely ${\bf E \cdot B = 0}$, were dropped. This is the
subject of the present work. A successful magnetospheric
configuration (solution) should be able to address the following
questions: \textbf{a)} Where does particle acceleration take place?
\textbf{b)} How close are these processes to a steady state?
\textbf{c)} Is the ensuing particle acceleration and radiation
emission consistent with the {\em Fermi} pulsar observations?

Unfortunately, while the ideal condition is unique, its
modifications are not. The evolution equations require a relation
between the current density $\mathbf{J}$ and the fields, i.e. a form
of Ohm's law, a relation  which is currently lacking.
\citet{meier2004} proposed a generalization of Ohm's law in plasmas
taking into account many effects (e.g. pressure, inertia, Hall).
However, in the present case we are in need of a macroscopic
expression for Ohm's law which would be also everywhere compatible
with the microphysics self-consistently, a procedure that might
involve Particle in Cell (PIC) techniques, well beyond the scope of
this paper. For this reason, the present work has a largely
exploratory character. A number of simple prescriptions are invoked
which lead to magnetospheric structures with values of $ {\bf
E}_{\vert \vert} \neq 0$. The global electric current and electric
field structures are then examined with emphasis on the
modifications of the corresponding FFE configuration.  We are
presently interested only in the broader aspects of this problem,
deferring more detailed calculations to future work. In particular,
the general compatibility of the $ {\bf E}_{\vert \vert}$ magnitude
with the values needed to produce (through pair production)
self-consistent charge and current densities from the plasma
microphysics will not be discussed in the present paper. The
contents of the present paper are as follows: In \S 2 we discuss the
different modifications in the expression for the electric current
that we apply in order to obtain non-Ideal solutions. In \S 3 we
present our results and the properties of the new solutions. In \S 4
we give particular emphasis to specialized solutions that obey
certain global conditions. Finally, in \S 5 we present our
conclusions.

\section{Non-Ideal Prescriptions}

In the FFE description of pulsal magnetospheres \citet{S2006} and
\citet{kc2009} solved numerically the time dependent Maxwell
equations
\begin{align}\label{maxeqb}
\frac{\partial \mathbf{B}}{\partial
t}&=-c\boldsymbol{\nabla}\times\mathbf{E}\\
\label{maxeqe} \frac{\partial \mathbf{E}}{\partial
t}&=c\boldsymbol{\nabla}\times\mathbf{B}-4\pi \mathbf{J}
\end{align}
under ideal force-free conditions
\begin{equation*}\label{IFFEcond}
\mathbf{E\cdot B}=0,\ \  \rho \mathbf{E+\frac{1}{c}J\times B}=0,
\end{equation*}
where $\rho=\mathbf{\nabla \cdot E}/(4\pi)$. The evolution of these
equations in time requires in addition an expression for the current
density $\mathbf{J}$ as a function of $\mathbf{E}$ and $\mathbf{B}$.
This is given by
\begin{equation}\label{IFFEJ}
    \mathbf{J}=c\rho\frac{\mathbf{E}\times\mathbf{B}}{B^2}+
    \frac{c}{4\pi}\frac{\mathbf{B\cdot \nabla\times B - E\cdot \nabla\times
    E}}{B^2}\mathbf{B}
\end{equation}
\citep{G1999}. The second term in Eq.~\eqref{IFFEJ} ensures that the
condition $\mathbf{E\perp B}$ (the ideal condition) is preserved
during the time evolution. However, this term introduces numerical
instabilities since it involves spatial derivatives of both fields.
Thus, \citet{S2006} implemented a scheme that evolves the fields
considering only the first term of Eq.~\eqref{IFFEJ}, and at the end
of each time step ``kills'' any developed electric field component
parallel to the magnetic field $(\mathbf{E_{\vert\vert}})$. In the
case of non-Ideal status, the $\mathbf{E}_\parallel =0$ condition
does not apply anymore and electric field components parallel to the
magnetic field are allowed. However, while in FFE the prescription
that determines the value of $\mathbf{E_{\vert\vert}}$ is unique (it
is equal to zero), there is no unique prescription in the non-Ideal
case. Below we enumerate the non-Ideal prescriptions that we
consider in the present study:

\noindent \textbf{(A)} The above implementation of the ideal
condition, hints at an easy generalization that leads to non-Ideal
solutions: One can evolve Eqs.~\eqref{maxeqb} \& \eqref{maxeqe},
using only the first term of the FFE current density
(Eq.~\ref{IFFEJ}), and at each time step keep only a certain
fraction $b$ of the $\mathbf{E_{\vert\vert}}$ developed during this
time instead of forcing it to be zero. In general the portion $b$ of
the remaining $\mathbf{E_{\vert\vert}}$ can be either the same
everywhere or variable (locally) depending on some other quantity
(e.g. $\rho, J$). As $b$ goes from 0 to 1 the corresponding solution
goes from FFE to vacuum. In this case an expression for the electric
current density is not given a priori, and $\mathbf{J}$ can be
obtained indirectly from the expression
\begin{equation}\label{ni11j}
    \mathbf{J}=\frac{1}{4\pi}\left(c\boldsymbol{\nabla}\times\mathbf{B}-
    \frac{\partial \mathbf{E}}{\partial t}\right)
\end{equation}

\noindent \textbf{(B)} Another way of controlling
$\mathbf{E_{\vert\vert}}$ is to introduce a finite conductivity
$\sigma$. In this case we replace the second term in
Eq.~\eqref{IFFEJ} by $\sigma \mathbf{E_{\vert\vert}}$  and the
current density reads
\begin{equation}\label{ni1j}
    \mathbf{J}=c\rho\frac{\mathbf{E}\times\mathbf{B}}{B^2}+
    \sigma \mathbf{E_{\vert\vert}}
\end{equation}
Note that Eq.~\eqref{ni1j} is related to but is not quite equivalent
to Ohm's law which is defined in the frame of the fluid. Others
\citep*{Lyutikov2003,lispitch2011} implemented a different version
closely related to Ohm's law. The problem is that the frame of the
fluid is not well defined in our problem. Moreover, it is not a
priori known neither that a single form of Ohm's law is applicable
in the entire magnetosphere, nor that the conductivity corresponding
to each such form of Ohm's law should be constant (e.g. in regions
where pair production takes place). This is why we chose a simpler
expression, which turns out to yield results qualitatively very
similar to other formulations based on Ohm's law. As is the case in
prescription (A) with the parameter $b$, here too, as $\sigma$
ranges from $\sigma=0$ to $\sigma\rightarrow\infty$ we obtain a
spectrum of solutions from the vacuum to the FFE, respectively. We
note that even though the vacuum and FFE solutions are unambiguously
defined limiting cases, there is an infinite number of paths
connecting them according to how $\sigma$ depends on the local
physical parameters of the problem ($\rho, E, B$ etc.).

\noindent \textbf{(C)} It has been argued
\citep{L1996,G2007,L2008_proc} that regions corresponding to
space-like currents $(J/\rho c > 1)$ are dissipative due to
instabilities related to counter streaming charge
flows.\footnote{Note that space-like currents $(J/\rho c > 1)$ are
formed exclusively by counter streaming charge flows.} Moreover,
the space-like current regions should trace the pair production
areas (i.e. dissipative areas). \citet{G2007} proposed a covariant
formulation for the current density that introduces dissipation
only in the space-like regions while the time-like regions remain
non-dissipative. The current density expression in the so called
Strong Field Electrodynamics (hereafter SFE) reads
\begin{equation}\label{SFEJ}
    \mathbf{J}=\frac{c\rho \mathbf{E\times B}+(c^2\rho^2+
\gamma^2\sigma^2E_{0}^2)^{1/2}(B_0\mathbf{B}+E_0\mathbf{E})}{B^2+E_0^2}
\end{equation}
where
\begin{equation}\label{SFEINV}
    B_0^2-E_0^2=\mathbf{B^2-E^2},\; B_0E_0=\mathbf{E\cdot B},\; E_0\geq 0\end{equation}
\begin{equation}\label{SFEgam}
    \gamma^2=\frac{B^2+E_0^2}{B_0^2+E_0^2}
\end{equation}
and $\sigma$ is a function of $E_0, B_0$ with dimensions of
conductivity (inverse time). The expression for $\mathbf{J}$
(Eq.~\ref{SFEJ}) produces either a space-like $(\sigma>0)$ or a null
$(\sigma=0)$ current. Note that this treatment precludes time-like
currents. Such currents do exist but they are only {\em effectively
time-like}: the current $J$, along with $E_{\parallel}$, fluctuates
continuously, so that while $ J$ is locally greater than (or equal
to) $\rho c$ its average value is less than that, i.e. $\langle J
\rangle < \rho c$. This behavior is captured by our code and is
manifest by the continuously changing direction of the parallel
component of the electric field along and against the magnetic field
direction \citep{G2008,G2011}.

In prescriptions (A) and (B) above, care must be taken {\em
numerically} so that the resulting value of $E_\perp$ (defined as
$|{\bf E}-{\bf E}_\parallel|$) be less than B. Violation of this
condition usually happens in the current sheet where the value of B
goes to zero. We note that the presence of the $E_0$ term in the
denominator of the SFE prescription (Eq.~\ref{SFEJ}) and of the
prescription of $\mathbf{J}$ used by \citet{lispitch2011} indeed
prevents the denominator from going to zero and alleviates this
issue in the current sheet regions. However, alleviation of this
problem does not guarantee that all treatments that include this
$E_0$ term converge to the same solution or that even produce the
same dissipation rates in the current sheet. Thus, the solutions and
in particular the dissipation rates of SFE and that of
\citet{lispitch2011} can be quite different.

All the simulations that are presented below have run in a cubic
computational box  20 times the size of the LC radius $R_{\rm LC}$,
i.e $[-10R_{\rm LC}\ldots 10R_{\rm LC}]^3$. Outside this area we
implemented a PML layer that allows us to follow the evolution of
the magnetosphere for several stellar rotations \citep{kc2009}. The
adopted grid cell size is $0.02R_{\rm LC}$, the stellar radius has
been considered at $r_{\star}=0.3R_{\rm LC}$, and all simulations
were run for 4 full stellar rotations to ensure that a steady state
has been achieved.

\section{Results}

We have run a series of simulations using prescription (B) that
cover a wide range of $\sigma$ values\footnote{We have found from
our simulations that the results of prescription (A) are very
similar to those of prescription (B). We therefore restrict the
discussion in the rest of this section to magnetospheres constructed
using prescription (B). An explicit reference to results under
prescription (A) is nonetheless made later on in this section in
relation to results shown in Fig. 5.}. In Fig.~\ref{fig01}a we
present the values of the Poynting flux $L$ integrated over the
surface of the star for the $a=0^{\circ}$ (aligned; red),
$a=45^{\circ}$ (green) and $a=90^{\circ}$ (perpendicular; blue)
versus $\sigma$ in log-linear scale. The two solid horizontal line
segments of the same color denote in each case ($a=0^{\circ},
45^{\circ}, 90^{\circ}$) the corresponding limiting values of the
Poynting flux for the FFE and the vacuum solutions, respectively.
The Poynting flux as a function of the inclination angle $a$ for the
vacuum and the FFE solutions reads
\begin{equation}\label{pf01}
    L=\begin{cases}
    \dfrac{2}{3}\dfrac{\mu^2 \Omega^4}{c^3}\sin^2a& \text{Vacuum}\\
    \\
    \dfrac{\mu^2 \Omega^4}{c^3}(1+\sin^2a)& \text{FFE}
    \end{cases}
\end{equation}
where $\mu, \Omega$ are the magnetic dipole moment and the
rotational frequency of the star, respectively. This means that for
a specific value of $a$ the corresponding value of the Poynting flux
(spin-down rate) of each non-Ideal modification should lie between
the two values of Eq.~\eqref{pf01}. Thus, Fig.~\ref{fig01}a shows
that for the perpendicular rotator (and any case with $a \ne 0$) the
ratio of the Poynting flux over that of the vacuum reaches near 1
for low values of $\sigma ~(\sigma/ \Omega \lsim 0.01$; for $a=0$
this is true only for $\sigma \rightarrow 0$). In Fig.~\ref{fig01}b
we plot for these simulations the corresponding dissipation power
\begin{equation}\label{energydissipation}
    \dot E_D=\int_{r_1<r<r_2}\mathbf{J}\cdot\mathbf{E}\ {\rm d}V=
    \frac{1}{4\pi}\int_{r_1<r<r_2}\sigma E_{\parallel}^2\ {\rm d}V
\end{equation}
taking place within the volume bounded by the spheres $r_1=0.3
R_{LC}$, i.e just above the star surface, and $r_2=2.5 R_{LC}$,
versus $\sigma$. We see that the energy losses due to dissipation
exhibit a maximum that occurs at intermediate values of $\sigma
~(\sigma/\Omega \simeq 1)$ for $a=90^{\circ}$ and at higher values
of $\sigma$ as the inclination angle $a$ decreases. The dissipation
power $\dot{E}_D$ goes to 0 as $\sigma$ goes either towards 0 or
$\infty$ since both the vacuum ($\sigma=0$) and FFE
($\sigma\rightarrow \infty$) regimes are (by definition)
dissipationless. We note that despite the fact that the Poynting
flux (on the stellar surface) varies significantly with $\sigma$,
the corresponding energy loss due to dissipation is limited and it
never exceeds the value $\simeq0.14 \mu^2 \Omega^4/c^3$. The surface
Poynting flux reflects the spin down rate while the dissipative
energy loss within the magnetosphere reflects the maximum power that
can be released as radiation. Moreover, when $\dot E_D$ is measured
as a fraction of the corresponding surface Poynting flux it is
always less than 22\% for $a= 0^{\circ}$, 15\% for $a= 45^{\circ}$
and 8\% for $a= 90^{\circ}$. These maximum fraction values occur at
lower $\sigma$ values than those that correspond to the maximum
absolute values of $\dot E_D$ (Fig.~\ref{fig01}b,c).

We note that all the values presented in Fig.~\ref{fig01} are the
corresponding average values over 1 stellar period (4$^\text{th}$
stellar rotation). Moreover, we note that the  values of $\dot E_D$
shown may be a little bit smaller than the true ones. This is
because our simulations introduce some numerical dissipation even
for the ideal-FFE solutions; for our resolution, this produces a
decrease in the Poynting flux, $\Delta L$, which at $r=3$ can reach
$\simeq 10\%$ of its value $L$ on the stellar surface. This means
that there is some energy loss that can not be measured by the
$\mathbf{J} \cdot \mathbf{E}$  expression for the values provided by
our solutions. Li et al. 2011 using higher resolution simulations
seem to obtain a corresponding decrease of $\approx 5\%$. Due to
this effect, the Poynting flux reduction with $r$ in the dissipative
simulations is always larger than its true value. This means that
the true $\dot E_D$ value is also a little higher than computed,
since the energy reservoir is a little larger. Nevertheless, the
error in $\dot E_D$ is less than $\Delta L /L$. We ran also
simulations with half grid cell size and simulations with the
stellar radius set at 0.45$R_{LC}$. The $\dot E_D$ values in these
cases were at most a few percent (mostly for small values of $a$) of
the corresponding $L$ values lower due to the finer description of
the stellar surface.

In Fig.~\ref{fig02} we present the magnetospheric structures that we
obtain on the poloidal plane $(\boldsymbol{\mu}, \mathbf{\Omega})$
using prescription (B) for $\sigma \simeq 24 \Omega$. Each row
corresponds to the structure at a different inclination angle
$a=(0^{\circ}, 45^{\circ}, 90^{\circ})$ as indicated in the figure.
The first column shows the modulus of the poloidal current $J_p$ in
color scale together with its streamlines. The second and third
columns show, in color scale, the charge density $\rho$ and
$E_{\parallel}$ respectively, together with the streamlines of the
poloidal magnetic field $\mathbf{B_p}$. We observe that for this
high $\sigma$ value the global structures are similar to those of
the FFE solutions \citep[see][]{ck2010}. Even for the $a=0^{\circ}$
case that has the most different value of the Poynting flux from
that of the FFE solution, the current sheet both along the equator
(outside the LC) and along the separatrices (inside the LC) does
survive despite the effects of non-zero resistivity. However, we do
observe magnetic field lines reconnecting gradually on the
equatorial current sheet beyond the LC. The main differences between
these solutions and the FFE ones are highlighted in the third column
where we plot the parallel electric field $E_{\parallel}$ developed
during the evolution of the simulation. For $a=0^{\circ}$ there is a
significant parallel electric field component along the separatrices
as well as over the polar caps\footnote{We note that the concepts of
the separatrix and the polar cap are less clearly in the case of
non-Ideal solutions since magnetic field lines close even beyond the
LC.}. The direction of the $\mathbf{E_{\parallel}}$ follows the
direction of the current $\mathbf{J}$. Thus,
$\mathbf{E_{\parallel}}$ within the return current region (the
separatrices and a small part of the polar cap near them) points
outward while in the rest of the polar cap region points inward.
There is also a weaker parallel component ${\bf E}_{||}$ along the
equatorial current sheet (not clearly shown in the color scale
employed). As $a$ increases, the inward pointing
$\mathbf{E_{\parallel}}$ component in the central part of the polar
cap becomes gradually offset. During this topological transformation
the branch of the polar cap area corresponding to the return current
becomes narrower while the other one becomes wider. At
$a=90^{\circ}$, $\mathbf{E_{\parallel}}$ is directed inwards in half
of the polar cap and outwards in the other half. This topological
behavior is similar to that of the poloidal current in the FFE
solutions. For $a\neq 0^o$ there is a clearly visible component of
$\mathbf{E_{\parallel}}$ in the closed field lines area as well as
along the equatorial current sheet outside the LC.

Figure \ref{fig03} shows the same plots as Fig.~\ref{fig02} and in
the same color scale but for $\sigma \simeq 1.5 \Omega$. Although
the non-Ideal effects have been amplified, the global topological
structure of the magnetosphere has not changed dramatically. We can
still distinguish traces of the current sheet even in the
$a=0^{\circ}$ case, whose Poynting flux is substantially reduced
from that of the FFE case. As noted above, the magnetic field lines
close now beyond the LC, due to the finite resistivity that allows
them to slip through the outflowing plasma. The parallel electric
field component $\mathbf{E_{\parallel}}$ is non-zero in the same
regions as those seen in the higher $\sigma-$values; its topology
also is not very different from that of these cases, but its maximum
value increases with decreasing $\sigma$. We note also that despite
the fact that the dissipative energy losses corresponding to the
various values of $a$ are quite different (see Fig.~\ref{fig01}a),
the corresponding maximum values of $E_{\parallel}$ are quite
similar.

We have also run many simulations implementing prescription (A) for
various values of the fraction $b$ of the non-zero parallel electric
field component $E_{\parallel}$. We have covered this way the entire
spectrum of solutions from the vacuum to the FFE one. The ensemble
of solutions that connects these two limits seems to be similar to
that of prescription (B). It seems that there are pairs of the
values of the fraction $b$ and the conductivity $\sigma$ that lead
to very similar (qualitatively) solutions. In Fig.~\ref{fig04} we
plot the poloidal magnetic field lines together with the parallel
electric field component $E_{\parallel}$ in color scale (similar to
the third column of Figs.~\ref{fig02} and \ref{fig03}) for $b=0.75$
which has the same Poynting flux with the simulation of prescription
(B) with $\sigma=1.5\Omega$. We see that the magnetospheric
structures of the two solutions are quite close (compare
Fig.~\ref{fig04} with the third column of Fig.~\ref{fig03}). Also,
our results are qualitatively very similar to those of
\citet{lispitch2011} who, as noted, implemented a much more
elaborate prescription for $\mathbf{J}$ based on Ohm's law.

Figure \ref{fig05} is similar to Fig.~\ref{fig01} but for
prescription (C) - i.e. SFE, assuming the value of $\sigma$ to be
constant. This prescription does not tend to the vacuum solution for
$\sigma=0$ but to a special solution with $J/\rho c=1$. As in
Fig.\ref{fig01} all quantities presented in this figure correspond
to average values over 1 stellar period ($4^{\text{th}}$ rotation).
The Poynting flux behaves accordingly, relaxing to $L \simeq 0.55
\mu^2 \Omega^4/c^3$ for $a=0^{\circ}$ (rather than zero), $L \simeq
1.0 \mu^2 \Omega^4/c^3$ for $a=45^{\circ}$ and $L \simeq 1.4 \mu^2
\Omega^4/c^3$ for $a=90^{\circ}$ as $\sigma \rightarrow 0$. These
values are 55\%, 65\% and 70\% of the corresponding FFE values,
respectively. For high $\sigma$ values ($\sigma \rightarrow \infty$)
the corresponding solutions have a Poynting flux close to that of
the FFE solutions.

The dissipation power in this case is given by
\begin{equation}\label{energydissipationsfe}
\begin{split}
    \dot E_D&=\int_{r_1<r<r_2}\mathbf{J}\cdot\mathbf{E}\ {\rm d}V\\
       &=\int_{r_1<r<r_2} E_{0}\sqrt{\rho^2 c^2+\gamma^2\sigma^2 E_0^2}\ {\rm d}V
\end{split}
\end{equation}
and, as expected, is non-zero even for $\sigma=0$. For $\sigma=0$ we
get the highest value for $\dot E_D$, either in absolute sense or as
a fraction of the Poynting flux. These reach 45\%, 25\% and 12\% for
$a=0^{\circ}, a=45^{\circ}, a=90^{\circ}$ respectively. These are
higher than the highest values we got using prescription (B). Their
increased magnitude is mostly due to the increase in the $\dot E_D$
of the SFE solutions on the current sheet outside the LC. We observe
also that the dissipation remains at high levels even for very high
values of the conductivity and only the $a=0^{\circ}$ case (which is
the most dissipative one) shows a clear tendency of decreasing $\dot
E_D$ with $\sigma$. For $\sigma \gg 1$ the dissipation energy inside
the LC decreases significantly while it remains almost constant
outside the LC. We have tried higher resolution simulations that
allow higher values of $\sigma$ {\em without any apparent decrease
of the dissipation on the current sheet}. This appears to confirm
Gruzinov's claim that SFE allows for non-zero dissipation on the
current sheet even for $\sigma\rightarrow \infty$.

Figure \ref{fig06} depicts the same quantities as Figs.~\ref{fig02},
\ref{fig03} but for prescription (C) i.e. SFE and for $\sigma \simeq
8\Omega $. We observe that the magnetospheric structures in this
case are very close to those of FFE (especially for $a>0^{\circ}$).
The magnetic field lines open not much farther than the LC and the
usual equatorial current sheet develops beyond that. In the third
column, which shows the parallel electric field, we see that in the
effectively time-like regions near the star the color is ``noisy''
as a result of the constant change of its sign during the evolution.
In the next section we are going to discuss possible implications of
these types of behavior in these regions. The regions with the
``calm'' (non noisy) electric fields correspond to space-like
currents. We discern these electric fields mostly along the
separatrix in the aligned rotator and along the undulating
equatorial current sheet in the non-aligned rotator. We note that
for $a=90^{\circ}$ the entire polar cap region consists of such
components. In Fig.~\ref{fig07} we present the  results for the same
prescription, i.e. (C) - SFE, but for $\sigma =0$. As we have
mentioned in \S 2, these cases are significantly different from the
vacuum solution. The magnetic field lines close well outside the LC
while we are able to discern traces of the equatorial current sheet.
We observe again the ``noisy'' regions near the star (this is the
way that SFE deals with time-like currents, namely as fluctuating
space-like ones). In the axisymmetric case ($a=0^{\circ}$) we see
that the value of $E_{\parallel}$ as well as the area over which it
occurs increase for the smaller values of $\sigma$. For $a >
0^{\circ}$ the value that $E_{\parallel}$ reaches near the polar cap
is not significantly higher that that of Fig.~\ref{fig06}. However,
the new interesting feature in the oblique rotators is the
development of high $E_{\parallel}$ in specific regions of the
closed field lines areas. These configurations do not imply high
dissipation since the corresponding currents are very weak there.

In order to reveal the on-average behavior of the effectively
time-like regions we considered the average values of the fields
within one stellar period. In Fig.~\ref{fignewsfeav} we plot for the
SFE ($\sigma=0$) the corresponding (average) parallel electric
components $\overline{E_{\parallel}}$ together with the magnetic
field lines (similar to the third columns in Figs.~\ref{fig06} and
\ref{fig07}). We see that the ``noisy'' behavior has disappeared. As
we discussed previously one can imagine the existence of hybrid
solutions (e.g. non ``noisy'' behavior in the closed field line
regions like that shown in Fig.~\ref{fignewsfeav}) and a ``noisy''
one in the open field line regions like that shown in
Fig.~\ref{fig06}.

Finally, we note that for all the solutions presented in this
section, the regions corresponding to high values of
$\mathbf{J}\cdot \mathbf{E}$ are traced mostly by the high values of
$J$ (e.g. near the current layers). However, as we observe in
Figs.~\ref{fig02}-\ref{fig03},~\ref{fig06}-\ref{fig07} there are
regions corresponding to high values of $E_{\parallel}$ that do not
lie near the current sheet. This implies that high energy particles
supporting the pulsar emission may be generated not near the regions
of high $\dot E_D$ but near those with high values of
$E_{\parallel}$ .

\section{Solutions with  $ J = \rho c$}

As argued above, the time evolution of the MHD equations requires a
prescription that relates the current density $\mathbf{J}$ and the
fields, given above by the different forms of Ohm's law. However,
given that any value of ${\bf E_{\parallel}}$ appropriate for
pulsars would quickly accelerate any charges to velocities close to
$c$, it would be reasonable to assume that the relation $J \simeq
\rho c$ will be attained between the current and charge densities.
Motivated by these considerations, rather than determine the current
$\mathbf{J}$ from its relation to $\mathbf{E_{\parallel}}$, we
determine it from its relation to the local charge density $\rho$.
Specifically, we focus our attention to solutions with $J/\rho c =
1$. This specific ratio is special because it discriminates between
magnetospheres consistent with only a single sign charge carrier
($J/\rho c \lsim 1$) and those that require carriers of both signs
of charges ($J/\rho c > 1$) to ensure the counter streaming required
in space-like current flows. Since each flow line with $J/\rho c >
1$ requires at some point a source of charges of the opposite sign,
it is considered that such flows by necessity involve the production
of pairs and their associated cascades.

One well-studied location for charge production in the pulsar
magnetosphere is the polar cap, where particle acceleration and pair
cascades are thought to take place.  If charges of either sign are
freely supplied by the neutron star surface, as is thought to be the
case in all but magnetar-strength fields \citep{medlai2010}, then a
maximum current of $J/\rho c = 1$ can flow outward above the polar
cap in Space-Charge Limited Flow (SCLF). In this case, an
accelerating electric field $E_{\parallel}$ is produced by a charge
deficit $\Delta\rho = (\rho - \rho_{GJ})$ that develops because the
GJ density is different from the actual charge flow along the field
lines \citep{arosch1979}. This electric field then supports a
current of primary electrons that radiates $\gamma$-rays, initiating
pair cascades.  The pairs efficiently screen the $E_{\parallel}$
above a pair formation front (PFF), by downward acceleration of
positrons. This charge density excess is small relative to the total
charge, i.e. $\Delta\rho/\rho = \epsilon \ll 1$ and the current that
flows into the magnetosphere above the PFF has $J /\rho c \simeq 1$
\citep{harmus2001}. Therefore the steady-state SCLF acceleration
models are only compatible with a small range ($J /\rho c \simeq 1$)
of the current to charge density ratio. \citet{bel2008} demonstrated
that no discharge occurs if $J /\rho c < 1$ and that the
$E_{\parallel}$ and the discharge generated when $J /\rho c > 1$ or
$J /\rho c < 0$ (implying that the current direction is opposite to
the one that corresponds to outward movement of the Goldreich-Julian
charges) are strongly time dependent. This has been confirmed by
models of time-dependent SCLF pair cascades \citep{timaro2011}.

With this emphasis on the ratio $J/\rho c =1$, we show in Figure
\ref{fig08} the values of $J/\rho c$ in color scale together with
the magnetic field lines on the poloidal plane for the parameters of
the simulations of prescription (B) given earlier in
Figs.~\ref{fig02}-\ref{fig03} and prescription (C) given in
Figs.~\ref{fig06}-\ref{fig07}. These plots can be directly compared
to the FFE magnetosphere of Fig.~\ref{fig00}.

The first two columns of Fig.~\ref{fig08} indicate that as $\sigma$
in prescription (B) ranges from large values ($\sigma = 24 \Omega$;
the FFE has strictly speaking $\sigma=\infty$) to 0 (vacuum) the
ratio $J/\rho c$ decreases tending to 0 for $\sigma\rightarrow
0$.\footnote{The oblique rotators have $J/\rho c$ ratios near 0 for
much lower values of $\sigma$ than those of the first column of
Fig.~\ref{fig08}.} This can be seen by inspection of
Eq.~\eqref{ni1j}, the first term of which corresponds always to a
time-like current, with a space-like current arising only through
the effects of the second term. Hence there must be some spatial
distribution of $\sigma$ values below which no space-like region
exists because the $\sigma E_{\parallel}$ term is small enough.

The next two columns of Fig.~\ref{fig08}, showing the $J/\rho c $
ratio of  Prescription (C) - SFE, are interesting because this
prescription leads to $J/\rho c =1 $ for $\sigma \rightarrow 0$,
while tending near the FFE values for $\sigma \rightarrow \infty$,
but only in the space-like current regions of the FFE solutions. The
time-like current regions of the FFE solutions are now only
``effectively time-like", as discussed by Gruzinov, resulting in the
``fluctuating" or ``noisy" behavior that is apparent there.

Suspecting that the ``fluctuating" (or ``noisy") behavior of the
effectively time-like current regions of SFE is due to the imposed
relation between $J$ and $\rho c$, we modified prescription (B) to
search for local values of $\sigma$ that would be consistent with $J
/\rho c = 1$. More specifically, we consider a $\sigma$ variable in
time and space so that the total $J$ provided by Eq.~\eqref{ni1j} is
everywhere and at every time step, equal to $\rho c$. However, we
found that there are regions where $E_{\parallel}$ goes to zero for
a current $J$ less than $\rho c$. In these regions the second term
in the expression for $J$, namely $\sigma E_{\parallel}$, tends to
the second term of the FFE current expression (Eq.~\ref{IFFEJ})
being unable to comply with the $J=\rho c$ constraint. These regions
demand extremely high values of $\sigma$ in order to achieve the
$J/\rho c=1$ relation due to the vanishing $E_{\parallel}$ values.
We have dealt with this numerically by setting a maximum value of
$\sigma$ ($=50\Omega$) that provided a realistic suppression of the
$E_{\parallel}$ in these regions.

Figure \ref{fig09} shows the results of such an approach. The first
three columns provide respectively the values of the poloidal
current $J_p$, charge density $\rho$ and parallel electric component
$E_{\parallel}$ and are similar to those presented in
Figs.~\ref{fig02}, \ref{fig03}, \ref{fig06}, \ref{fig07}, while the
fourth column shows the values of the $J/\rho c$ ratio in color
scale together with the poloidal magnetic field lines. The
corresponding $L$ and $\dot E_D$ values as well as their ratios
$\dot E_D/L$ have been denoted in Fig.~\ref{fig01} by the horizontal
dashed lines. Not surprisingly, these solutions are similar to those
of prescription (C) - SFE in the $\sigma\rightarrow 0$ regime (see
Fig.~\ref{fig07}). The essential difference between these two
approaches is that, in distinction with SFE, this approach can
handle time-like currents without the need - like SFE -  to do so
only on average (see blue and purple color regions in the fourth
column).

To allow our solutions even larger flexibility in attaining $J =
\rho c$ we have looked for prescriptions that divorce $\mathbf{J}$
from ${\bf E_{\parallel}}$, replacing the second term of
Eq.~\eqref{ni1j}, with a term  of the form
\begin{equation}\label{jparallelb}
    \mathbf{J_{\parallel}}=f\;\mathbf{B},
\end{equation}
where $f$ is a scalar quantity having a local value that ensures the
local value of the ratio $J/\rho c$ to be equal to 1. The difference
from the previous approach and that of prescription (B) is that this
term is not related anymore to the value of $E_{\parallel}$, other
than that it is directed along (or against for $f <0$) the direction
of the magnetic field. The sign of $f$ is determined so that the
direction of $\mathbf{J_{\parallel}}$ coincides with that of
$\mathbf{E_{\parallel}}$. The results of these simulations are shown
in Fig.~\ref{fig10} the panels of which are similar to those of
Fig.~\ref{fig09}. The global structure in this case is very similar
to that of the simulation presented in Fig.~\ref{fig09} except for
the time-like regions (especially those near the star) that now
exhibit the ``noisy" behavior of the SFE simulations. However, the
ratio $J/\rho c$ is now everywhere equal to 1. Our solution has
therefore the required property ($J/\rho c=1$) at the expense of
producing a ``noisy" behavior indicative of time variations at these
limited spatial scales.

This oscillatory behavior (both here and in the SFE cases) is a
rather generic feature that occurs in the time-like current regions
for values of $J/\rho c$ larger than those needed to ``kill" the
local $\mathbf{E}_{\parallel}$ { implied by the field evolution
equations; both within the prescription of Eq.~\eqref{jparallelb}
and within SFE, at these higher values,  $J$ and $E_{\parallel}$ end
up pointing in different directions and an oscillatory behavior
ensues}. Smaller values of $J/\rho c$, cannot drive ${\bf
E_{\parallel}}$ to zero and allow the magnetosphere to attain steady
state, leaving a non-oscillatory component of
$\mathbf{E}_{\parallel}$ along the direction of
$\mathbf{J_{\parallel}}$. Viewed this way, the ideal condition ${\bf
E \cdot B = 0}$ is one that requires a local delicate balance
between its parameters (e.g. $J, \rho$). If, for whatever reason or
necessity, this balance breaks, the magnetosphere develops locally
an oscillatory behavior. The robustness and the true evolution of
this behavior will be determined by the microphysical plasma
properties as they mesh within the global framework of the
magnetosphere electrodynamics.

The conclusion of this exercise is that solutions with a specific
value of $J$, for example  $ J = \rho c$ everywhere, can be
implemented with a sufficiently general expression for
$\mathbf{J_{\parallel}}$, provided that one is willing to accept
locally time varying solutions.

Note that in this prescription, as well as in SFE, the current is
forced to follow the relation $J = \rho c$ or $J > \rho c$ (for
$\sigma\ne 0$ in SFE) even in the closed field line regions
producing even there this ``noisy" (fluctuating) behavior. Although
applying this $J - \rho$ prescription there is not forbidden a
priori, it would be reasonable to consider that this condition is
actually not valid in the closed field lines and that a different
condition should be used there. However, the implementation of a
different prescription for these regions remains cumbersome because
it demands that we determine whether the magnetic field lines at
each grid point are actually open or closed and then apply for them
this different condition. Nevertheless, even if we were to enforce a
prescription that removes the fluctuating (noisy) behavior in the
closed field line regions, this should not affect the similar
behavior on the open field line regions.

\section{Discussion and Conclusions}

Up to now, the pulsar radiation problem has been studied considering
either the vacuum \citep{deutsch1955} or the FFE solutions
\citep{ckf99,S2006,kc2009} for the structure of the underlying
magnetosphere. The vacuum solutions have analytical expressions
while the FFE solutions have only been studied numerically in the
last decade. However, neither solution is compatible with the
observed radiation; the former because it is devoid of particles and
the latter because it precludes the presence of any ${\bf
E_{\parallel}}$. The physically acceptable solutions have to lie
somewhere in the middle. As of now, there are no known solutions
that incorporate self-consistently the global magnetospheric
structure along with the microphysics of particle acceleration and
radiation emission. The goal of the present work has been to explore
the properties of pulsar magnetospheres under a variety of
prescriptions for the macroscopic properties of the underlying
microphysics in anticipation of a more detailed future treatment of
these processes. It should be noted that these processes lie outside
the purview of the equations used to evolve the electric and
magnetic fields. Nonetheless, for most models examined the
dissipation power does not exceed 10-20\% of the spin-down one; this
value is consistent with the observed radiative efficiencies of the
millisecond pulsars \citep{abdoetalsci2009}, that are the ones
reaching the highest efficiencies of all pulsars, suggesting that
energetically, realistic magnetospheres may be well approximated by
our models.

In order to explore the non-Ideal ${\bf E \cdot B \ne 0}$ regime we
considered several prescriptions discussed in \S 3, 4 and analyzed
their properties in detail (Figures 2 - 9). Despite their intrinsic
interest, the solutions discussed above still need to be examined
more closely for consistency with the microphysics responsible for
the set up and closure of the corresponding magnetospheric circuits,
which are characterized by a considerable range in the distribution
of the ratio $J/\rho c$. On the other hand, there are studies
arguing  that it is essential that the ratio $J/\rho c$ be near 1 in
order for steady-state pair cascades to be generated near the polar
cap. Besides that, solutions corresponding to values $J/\rho c<1$
can be considered as the range of configurations without pair
production, since such solutions can be supported by charge
separated flows.

With the value $J/\rho c = 1$ given this particular significance, we
have pursued a search for solutions that are driven by adherence to
this condition in as broad a spatial range as possible. As noted in
the previous section we did so by searching for solutions of
variable conductivity $\sigma$, or by employing expressions for the
parallel current unconstrained by the values of the parallel
electric field. We found that arbitrary values of $\sigma$, while
allowing $J/\rho c = 1$ in most areas, in certain others can drive
$\mathbf{E}_{\parallel}$ to zero sufficiently fast to restrict $J$
to values $J < \rho c $. As noted in the previous section, a more
general expression for $J$ not related to $\mathbf{E}_{\parallel}$
can produce $J/\rho c = 1$ everywhere but only at the expense of
fluctuations in regions where the value of $J/\rho c$ is larger than
that necessary to null  the local $\mathbf{E}_{\parallel}$ implied
by the field evolution equations. These results indicate the
importance of the value of the ratio $J/\rho c$ vis-\'a-vis the
presence of pair cascades and steady-state emission in pulsar
magnetospheres.

The actual behavior of the magnetosphere is likely to be more
complicated than that described by the current density prescriptions
provided in this work (Eqs.~\ref{ni1j},~\ref{SFEJ} and
\ref{jparallelb}). Physics beyond those of the magnetic and electric
field evolution equations will introduce novel time scales that will
determine the response of the current density to the parallel
electric field component and vice versa. The implications of such an
oscillatory behavior which involves an electric field that
alternates direction, may in fact provide an account of the pulsar
coherent radio emission, as such a behavior is the essential
ingredient of cyclotron maser emission
\citep{melrose1978,lmjl2005,melrose2006,luomel2008}. To the best of
our knowledge, the above argument is the first ever to suggest a
relation between the global properties of pulsar magnetospheres and
their coherent radio emission.

These more general models of non-ideal magnetospheres discussed
above are possibly related to the behavior and properties of
intermittent pulsars, which may, if more of them are discovered,
provide meaningful constraints on their radiation mechanism and
magnetospheric dynamics. Intermittent pulsars are pulsars found in
``on" and ``off" radio states \citep{lyne2009}. The interesting
aspect of these objects is that they exhibit different spin-down
rates in their ``on" and ``off" states; even more interesting is the
fact that they apparently ``remember" the spin-down rate they had in
the previous cycle, as they return to the same spin-down rate when
switching from ``on" to ``off" and vice versa. The ratios of the
``on" to ``off" spin-down rates are 1.5 -- 1.7 for pulsars
PSR~B1931+24 \citep{krameretal2006} and PSR~J1832+0029
\citep{lorimeretal2006,lyne2009}, while it is 2.5 for the more
recently observed PSR~J1841-0500 \citep{cetal2011}. Interestingly
these values are well below the minimum ratio 3 between the FFE and
the vacuum spin-down rates. Based on this observation,
\citet{lispitch2011} argued that the ``on'' state can be the FFE
solution while the ``off'' state simply a resistive solution
corresponding to a certain value of $\sigma$.

\begin{figure}
  \centering
  \includegraphics[width=5cm]{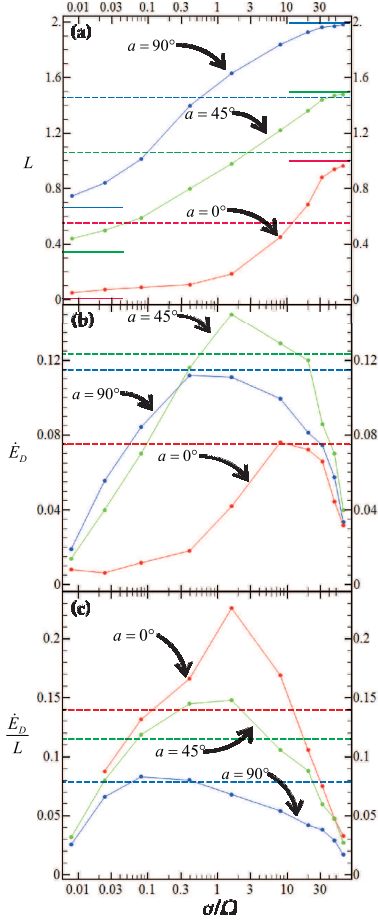}\\
  \caption{\textbf{(a)} The Poynting fluxes $L$ (measured on the surface of the star)
  as a function of $\sigma$ in log-linear scale for prescription (B).
  Red, green and blue colors correspond to $a=0^{\circ}$ (aligned), $a=45^{\circ}$ and $a=90^{\circ}$
  perpendicular rotator, respectively. The horizontal solid line elements denote the $L$ values
  corresponding to the vacuum (lower value) and the FFE (higher value) solutions.
  Note that the $L$ value for $a=0^{\circ}$ is similar to the vacuum one
  for $\sigma \lesssim 0.3 \Omega$ ($\Omega$ is the angular frequency of the star) and
  it reaches that of the FFE solution only for much higher $\sigma$ values;
  \textbf{(b)} The dissipation energy rate $\dot E_D$ integrated over the volume bounded by
  radii $r_1=r_{\star}=0.3R_{\text{LC}}$ and $r_2=2.5R_{\text{LC}}$ as a function of
  $\sigma$;
  \textbf{(c)} The fraction $\dot E_{D}/L$ never exceeds the values 10\%-20\%, while for
  $\sigma\rightarrow 0$ and $\sigma\rightarrow \infty$ it goes towards 0.
  The dashed horizontal lines in all three panels denote the values
  corresponding to simulations with $J/\rho c=1$ (see \S 4, 5).}\label{fig01}
\end{figure}

\begin{figure*}
  \centering
  \includegraphics[width=15cm]{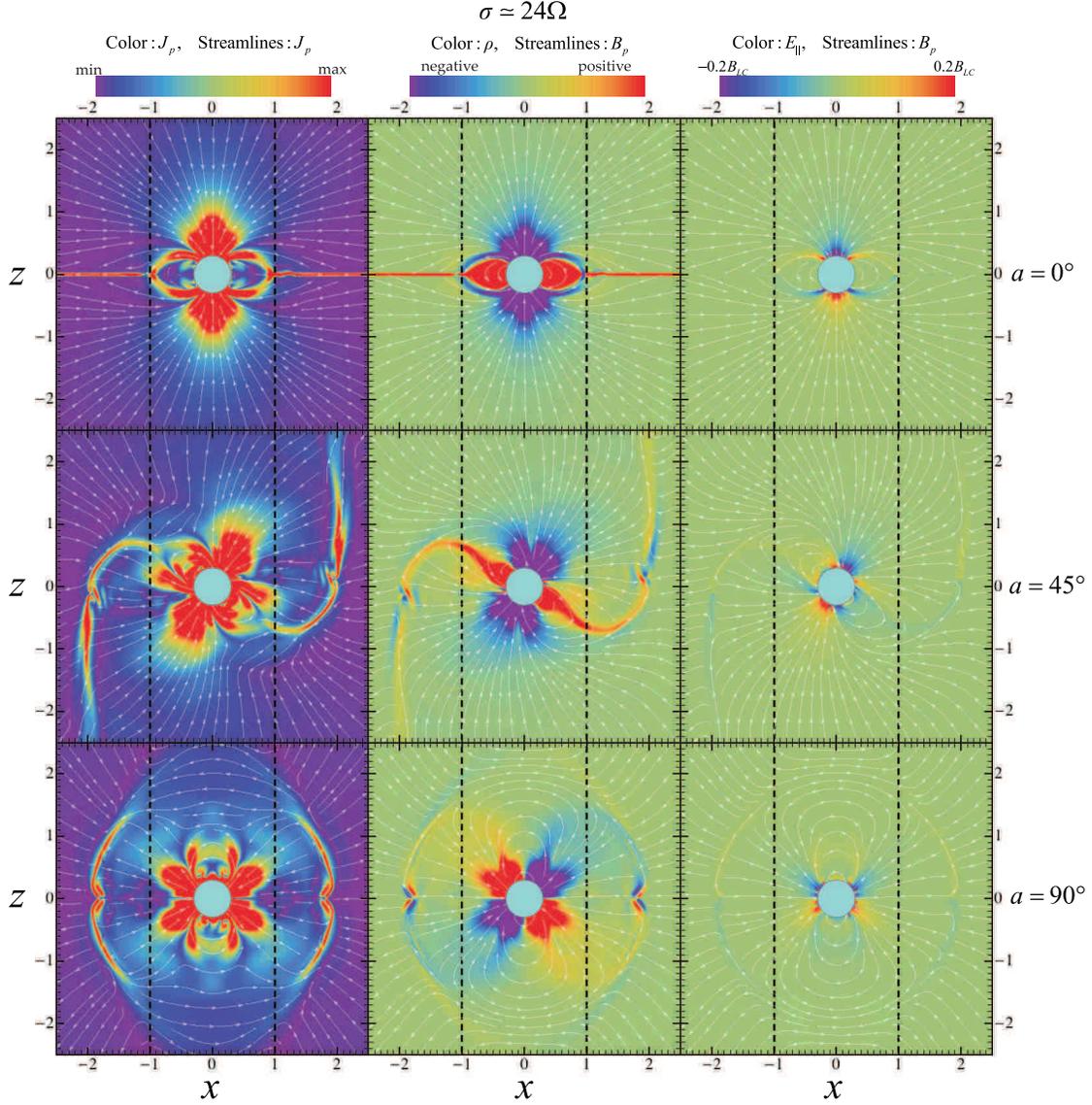}\\
  \caption{The non-Ideal magnetospheric solutions in the poloidal plane
  $(\boldsymbol{\mu}, \mathbf{\Omega})$ for prescription (B) and for
  a high $\sigma$ value ($\sigma=24\Omega$).
  Each row shows the solutions corresponding to the indicated inclination angles $a$.
  The left-hand column shows the poloidal current modulus (in color scale)
  together with the streamlines of the poloidal current.
  The middle column shows the charge density (in color scale) together with the field lines of the
  poloidal magnetic field. The color ranges purple-green and green-red indicate negative and positive
  charge density, respectively. The right-hand column shows the parallel electric field component
  $E_{\parallel}$ (in color scale) together with the lines of the poloidal magnetic field.
  The color ranges purple-green and green-red indicate antiparallel and
  parallel directions of $E_{\parallel}$ (relative to the magnetic field) respectively. Note
  that the color representation for $E_{\parallel}$ saturates beyond the absolute value
  $0.2B_{LC}/c$ where $B_{LC}$ is the value of the stellar magnetic dipole field at
  the distance $R_{\text{LC}}$. The structure shown in the first two columns
  is quite similar to that of the FFE solutions (the main difference from the
  FFE solution can be seen in the third column).}\label{fig02}
\end{figure*}

\begin{figure*}
  \centering
 \includegraphics[width=15cm]{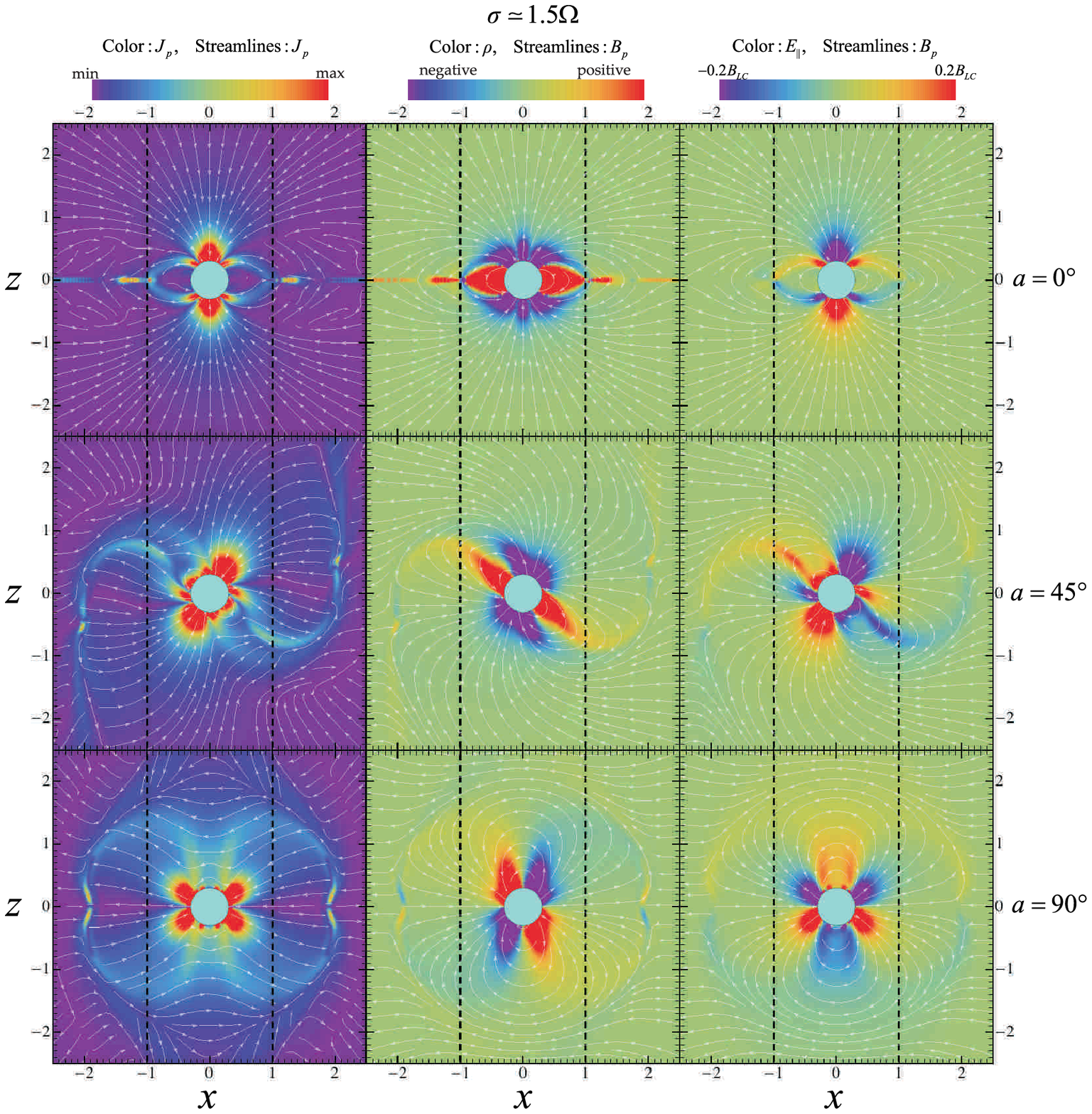}\\
  \caption{Similar to Fig.~\ref{fig02} but for a much lower $\sigma$ value ($\sigma=1.5\Omega$).
  The main magnetospheric features such as separatrices and current sheets can still be observed,
  even though they now appear much weaker. The magnetic field lines close well beyond
  the LC and the parallel electric field components reach higher values than those of
  Fig.~\ref{fig02}}\label{fig03}
\end{figure*}

\begin{figure}
  \centering
  \includegraphics[width=8cm]{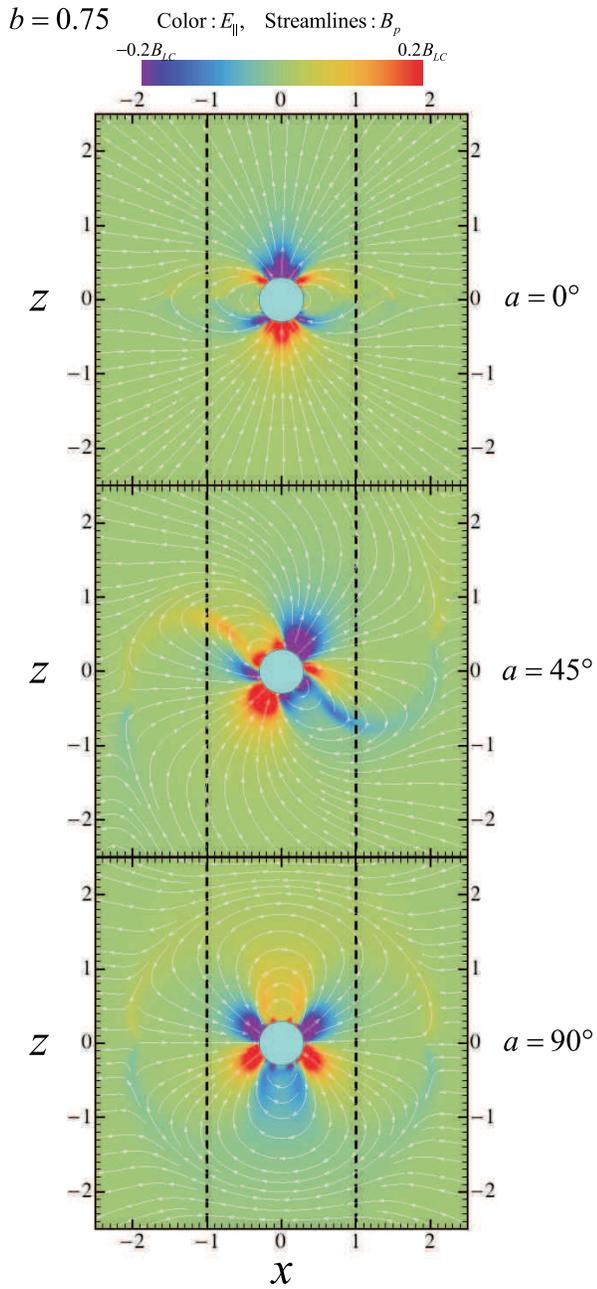}\\
  \caption{Similar to the third columns of Figs.~\ref{fig02}, \ref{fig03} but
  for the prescription (A) and for the $b=0.75$ value. For this value of $b$
  the corresponding $L$ values are the same to those of the simulations
  presented in Fig.~\ref{fig03}. The structure of these solutions seems very similar
  (qualitatively) to those of Fig.~\ref{fig03}.}\label{fig04}
\end{figure}

\begin{figure}
  \centering
  \includegraphics[width=5cm]{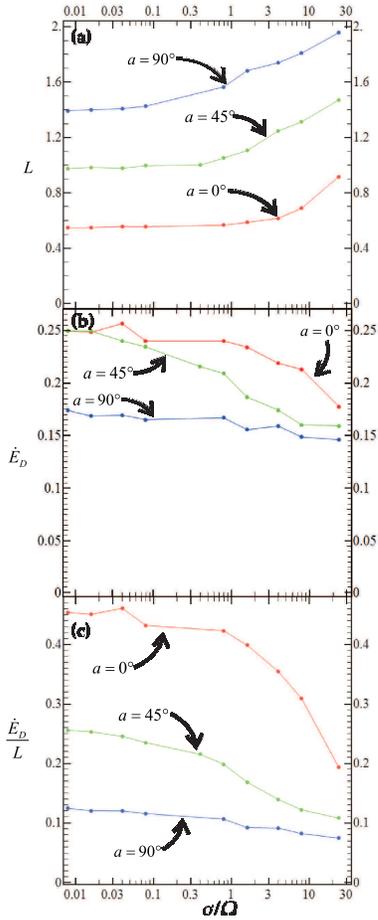}\\
  \caption{Similar to Fig.~\ref{fig01} but for prescription (C) - SFE.
  This prescription does not cover the entire spectrum of solutions
  between the vacuum and FFE ones since for $\sigma\rightarrow 0$ we still
  get a dissipative configuration rather than the vacuum solution. However, as
  $\sigma\rightarrow \infty$, the corresponding Poynting flux values tend to
  those of the FFE solutions. The dissipative energy loss rate $\dot E_D$
  exhibits a maximum value for $\sigma\rightarrow 0$ and remains non-zero
  even for the high $\sigma$ values due to the persistent dissipation on the
  current layer outside the LC. This result seems to confirm Gruzinov who argued
  that the SFE solutions for $\sigma\rightarrow \infty$ present finite dissipation
  on the current layer.}\label{fig05}
\end{figure}

\begin{figure*}
  \centering
  \includegraphics[width=15cm]{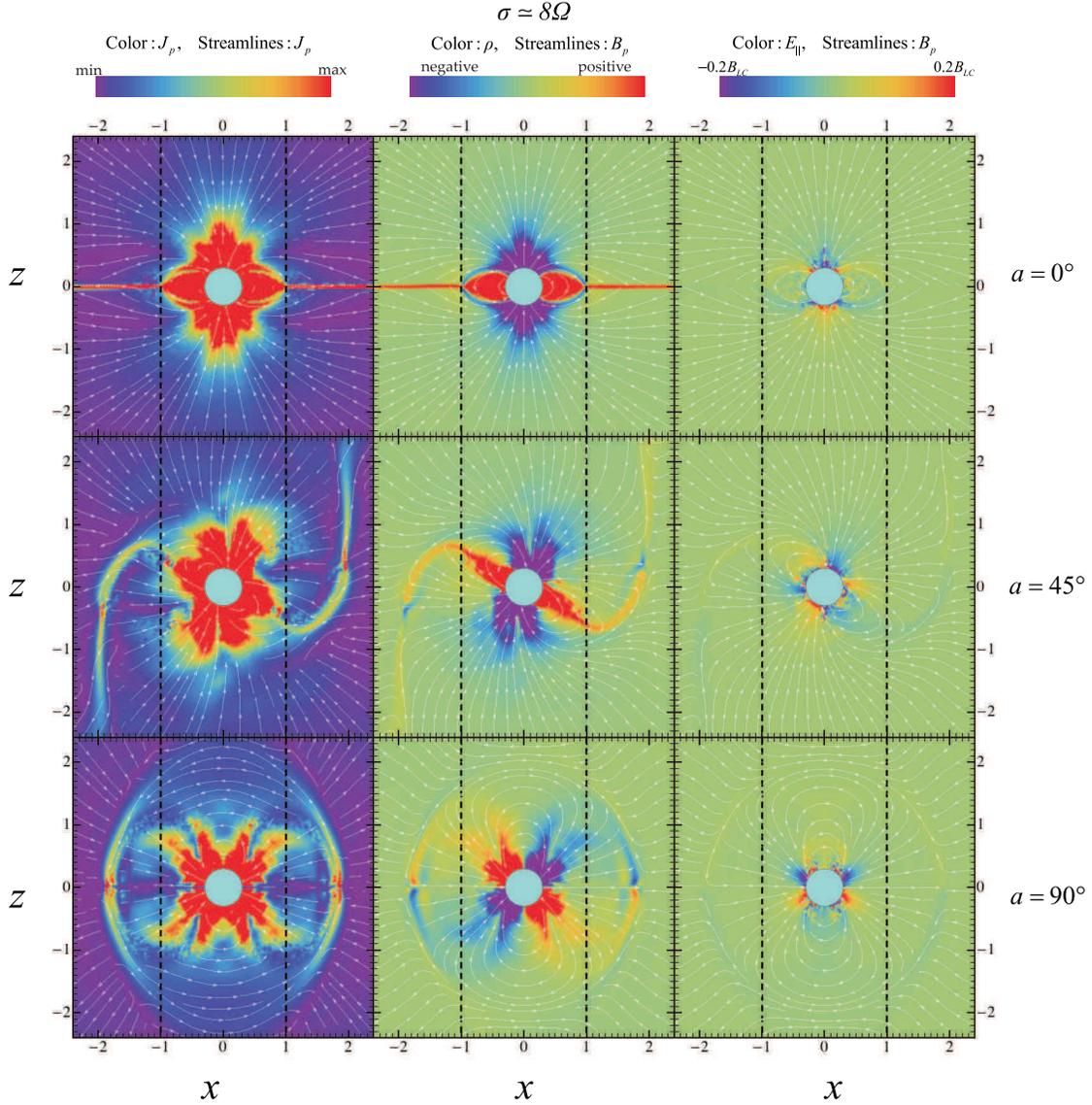}\\
  \caption{Similar to Figs.~\ref{fig01}, \ref{fig02} but for prescription (C) - SFE and for a
  high $\sigma$ value ($\sigma=8 \Omega$). The Poynting flux of these solutions tend to those of the
  FFE ones. Nevertheless, we still see magnetic field lines closing well outside the LC
  (especially for $a=0^{\circ}$). The electric current in the SFE prescription is
  everywhere space-like ($J/\rho c>1$). However, there are regions that are effectively
  time-like (as defined in the text). These are the noisy regions in the polar cap
  vicinity (for $a\neq 90^{\circ}$) seen in the third
  column. Their noisy character is due to the continuously changing
  direction of the electric field.
  }\label{fig06}
\end{figure*}

\begin{figure*}
  \centering
  \includegraphics[width=15cm]{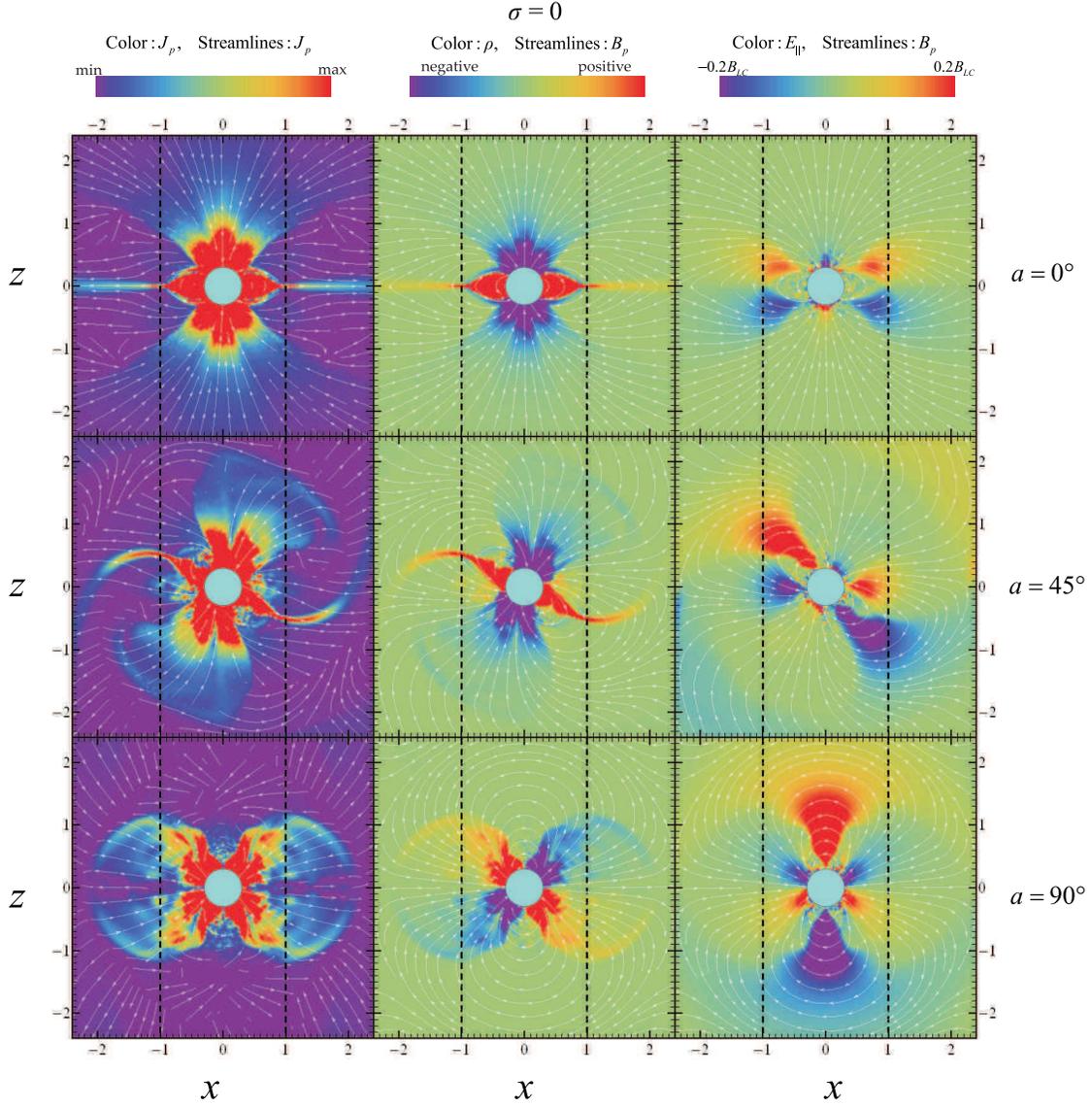}\\
  \caption{Similar to Fig.~\ref{fig06} but for $\sigma=0$ value.
  The modulus of the parallel electric field components increases along with the area
  of their influence. The closed field line regions of the oblique rotators
  show significant parallel electric field components; however, the dissipation
  there is insignificant because the value of the product
  $\mathbf{J}\cdot \mathbf{E}$ is small.}\label{fig07}
\end{figure*}

\begin{figure}
  \centering
  \includegraphics[width=6cm]{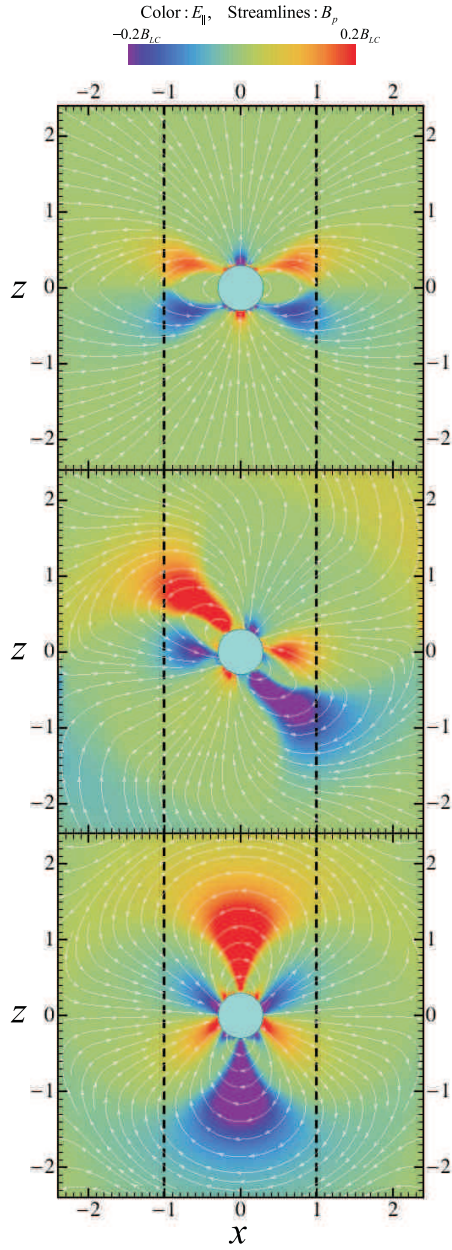}\\
  \caption{The third column of Fig.~\ref{fig07} considering the average values
  of the fields within a stellar period. The ``noisy'' behavior shown in
  Fig.~\ref{fig07} is now disappeared. These regions reveal their effectively
  time-like behavior.
  }\label{fignewsfeav}
\end{figure}

\begin{figure*}
  \centering
  \includegraphics[width=\textwidth]{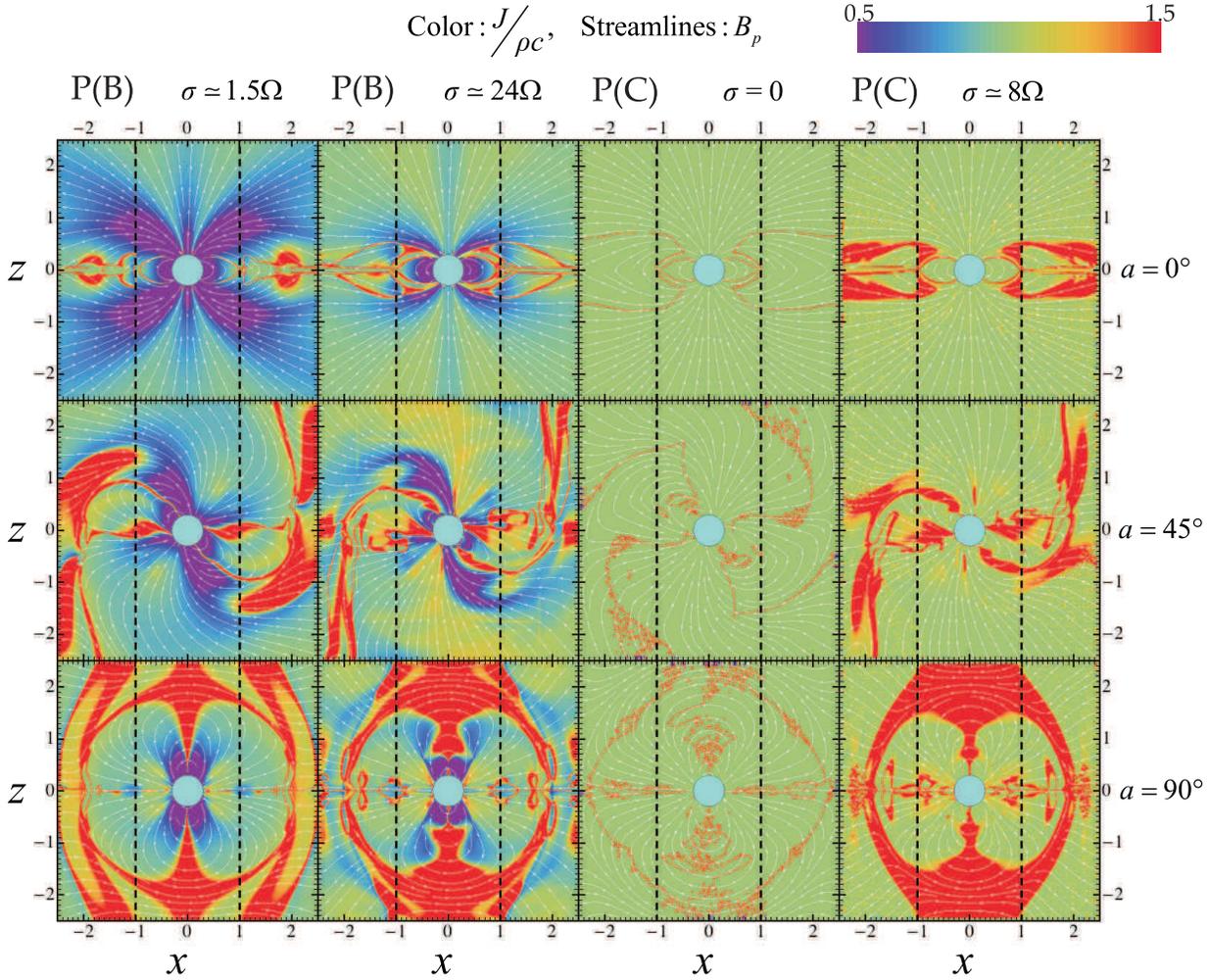}\\
  \caption{The ratio $J/\rho c$ (in color scale) together with the poloidal magnetic
  field lines for the solutions presented in Figs.~\ref{fig02}, \ref{fig03},
  \ref{fig06}, \ref{fig07} as indicated in the Figure. The ratio $J/\rho c$
  in prescription (B) decreases in general as $\sigma$ decreases and goes towards 0
  for $\sigma\rightarrow 0$. In prescription (C) - SFE the ratio $J/\rho c$
  goes to 1 (null current) as $\sigma\rightarrow 0$ while it is higher than 1
  for $\sigma>0$. However, in both these cases there are regions where the current
  has on average time-like ($J/\rho c <1$) behavior (effectively time-like;
  see Fig.~\ref{fignewsfeav}).}\label{fig08}
\end{figure*}

\begin{figure*}
  \centering
  \includegraphics[width=\textwidth]{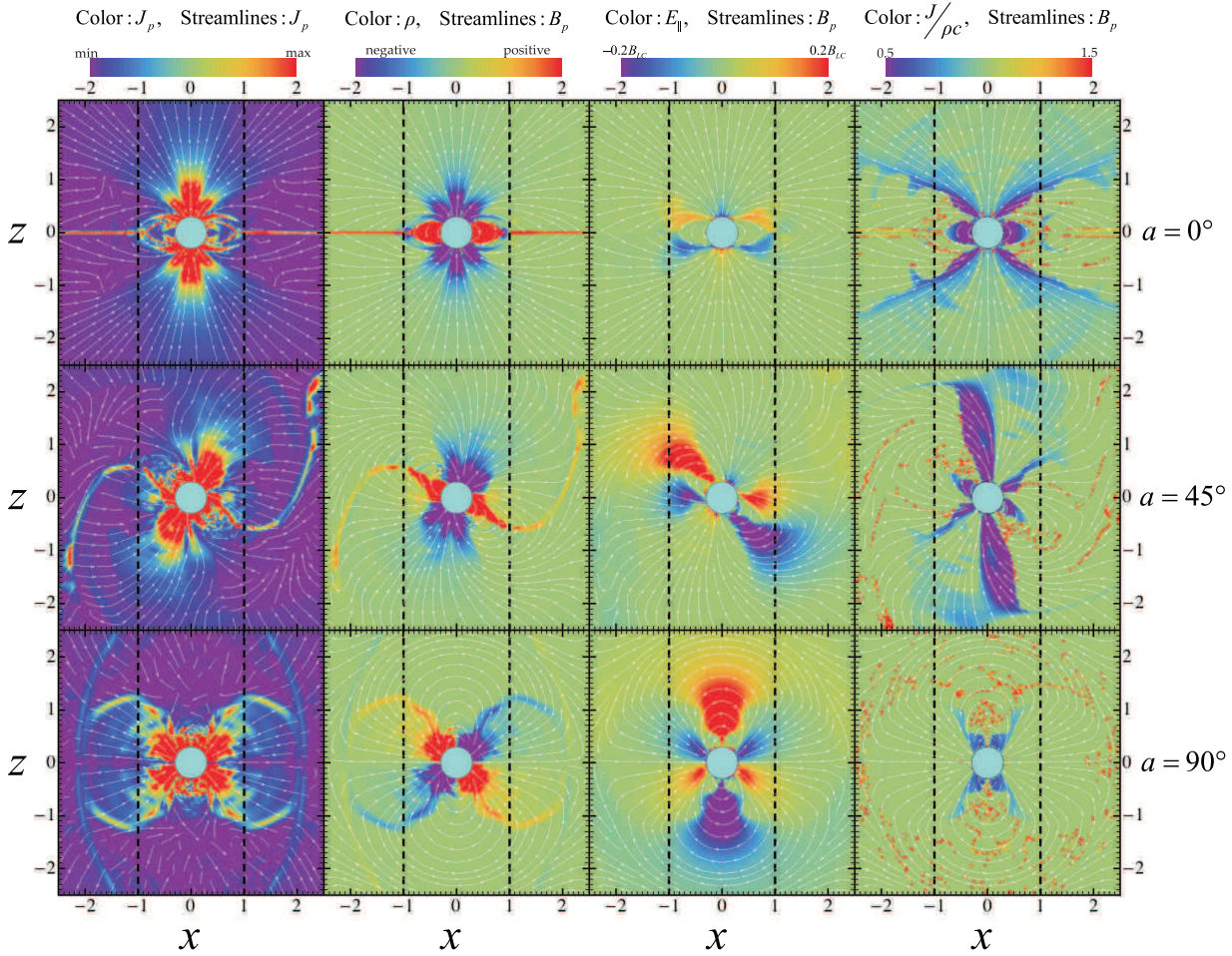}\\
  \caption{The solutions employing prescription (B) with the $\sigma-$values adapted locally to
  achieve $J/\rho c$ as close to 1 as possible. The first three columns are similar to those of
  Figs.~\ref{fig02}, \ref{fig03}, \ref{fig06}, \ref{fig07} while the fourth column
  shows the corresponding ratio $J/\rho c$ in color scale. The main features
  (e.g. current sheets, separatrices) of the FFE solutions are still discernible.
  We observe parallel electric field components across the polar caps and along the
  separatrices. For the oblique rotators parallel electric field components exist also
  in parts of the closed field line regions. However, the latter regions show low
  dissipation energy rate $\dot E_D$ since the corresponding product
  $\mathbf{J}\cdot \mathbf{E}$ is small. In the fourth column we observe regions with
  $J/\rho c<1$. In these regions a current density $J$ smaller than the value $\rho c$
  is enough so that the corresponding parallel electric field component
  vanishes. This kind of solutions are the upper limit of the solutions that can be considered that
  are supported by charge separated flows.}\label{fig09}
\end{figure*}

\begin{figure*}
  \centering
  \includegraphics[width=\textwidth]{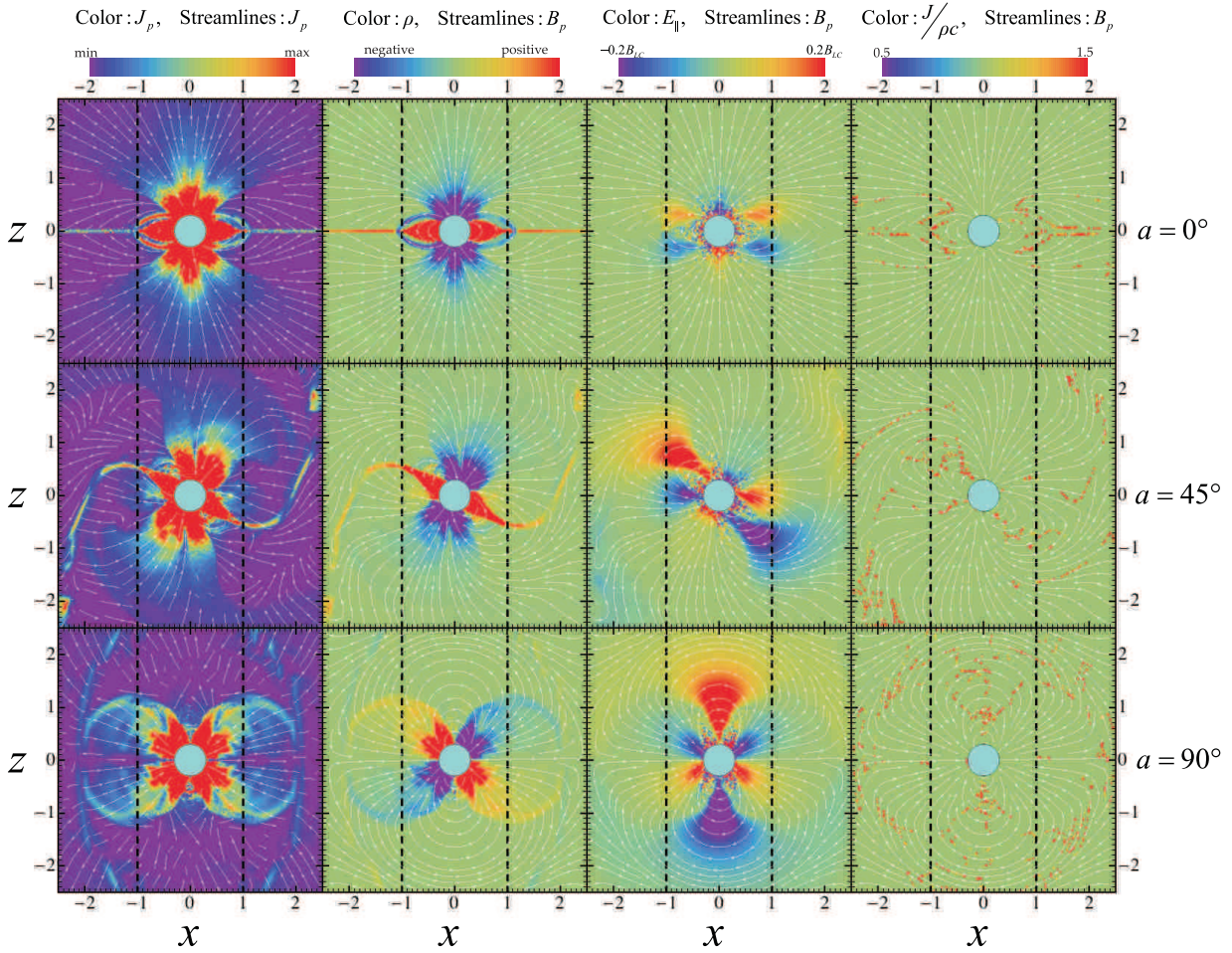}\\
  \caption{Similar to Fig.~\ref{fig09} but with  the parallel current density independent
  of the modulus of ${\bf E_{\parallel}}$ (see Eq.~\ref{jparallelb}). As shown in the fourth column,
  this prescription achieves $J/\rho c=1$ in the entire magnetosphere. This
  leads to an oscillating behavior for ${\bf E_{\parallel}}$ in the time-like current regions
  of the simulation presented in Fig.~\ref{fig09}, similar to that of SFE (Fig.~\ref{fig07}).
  We note that this effect is more general and appears wherever the ratio $J/\rho c$
  is enforced to have values above those that are sufficient to cancel out the parallel
  electric component. This behavior is the result of the mutual response between
  the electric field and the current density and may be crucial for the coherent radio emission.}\label{fig10}
\end{figure*}

An alternative view is to identify the ``on" state with
magnetosperes with $J / \rho c > 1$ and the ``off" one with those of
$J / \rho c < 1$, given the necessity of the presence of pairs
(accompanied by the radio emission) in the former and their likely
absence (and also of radiation) in the latter. In fact, the ratio of
spin-down rates between the FFE and the $J/\rho c = 1^-$
magnetospheres is within $1.4-1.8$ (see Figs.~\ref{fig01} and
\ref{fig05}), with the higher values corresponding to the lower
values of $a$. This would be consistent with the observations of the
first two intermittent pulsars. However, this is not consistent with
the spin-down ratio of the third intermittent pulsar whose ``on''
and ``off'' states may be magnetospheres with different values of
$\sigma$. The discovery of a larger sample of intermittent pulsars
(especially if one could also put some limits on their inclination
angles) would be extremely useful in constraining magnetospheric
models. For example if the ratio of the ``on/off" spin-down rate was
found not to exceed the value 3, one would have to exclude the
vacuum solution being the ``off" state, since for sufficiently small
$a$ this ratio becomes arbitrarily large.

The solutions we presented in this work, besides their intrinsic
interest, can serve to produce model pulsar $\gamma-$ray light
curves in a way similar to that of
\citet{ck2010,BS10b,venhargui2009,hardingetal2011}. Such models will
provide a means of connecting the structure and physics of
magnetospheres to observations, since they constrain the pulsar
magnetospheric geometry as well as the regions of anticipated
particle acceleration and even the values of the accelerating
electric fields. These will therefore provide independent tests of
these models for real pulsars.

The next level of this study should include microphysics at a level
sufficient to allow some feedback on the global solutions i.e. the
production of pairs whose effects are included in the computation of
the local $E_{\parallel}$ self--consistently. It will be of interest
to examine whether such solutions are consistent with the charge
separated flows or whether they require an inherently time dependent
magnetosphere, a feature that could be tested observationally. The
time dependence should lead to oscillations of the pulsar emission
with frequency of order $\nu \gsim \Omega/2\pi$ for emission near
the LC and $\nu \gg \Omega/2\pi$ for emission near the polar cap. We
have searched and found not these oscillations in the X-ray domain
of several pulsar light curves. Either they were not present at the
objects we searched or they lose their coherence at these much lower
energies. Nonetheless, we intend to study the possibility of the
presence of such oscillation in a future work.

\acknowledgments We would like to thank the anonymous referee for
constructive comments that helped us to improve this paper.

\clearpage

\clearpage

\clearpage

\clearpage

\clearpage

\clearpage

\clearpage

\clearpage

\clearpage

\clearpage

\clearpage

\clearpage

\clearpage

\bibliographystyle{apj}
\bibliography{references_all}

\label{lastpage}

\end{document}